\documentstyle[12pt]{article}
\input{eepic.sty}

\def\a{\alpha}
\def\b{\beta}

\def\l{\lambda}

\def\de{\delta}
\def\t{\theta}
\def\ep{\epsilon}

\def\P{\varphi}

\def\half{{1\over 2}}

\begin{document}

\title{Boundary Flows in general Coset Theories}

\author{Changrim Ahn
\footnote{Permanent Address: Department of Physics,
Ewha Womans University, Seoul 120-750, Korea} \\
\small Institut f\"ur Theoretishe Physik, Freie Universit\"at Berlin\\
\small Arnimallee 14, Berlin D-14195, Germany
\\
\small and
\\
Chaiho Rim\\
\small Department of Physics\\
\small Chonbuk National University, Chonju 561-756, Korea
}

\date{}

\begin{titlepage}

\maketitle

\begin{abstract}

In this paper we study the boundary effects for 
off-critical integrable field theories which have close 
analogs with integrable lattice models.
Our models are the $SU(2)_{k}\otimes SU(2)_{l}/SU(2)_{k+l}$ coset
conformal field theories perturbed by integrable boundary 
and bulk operators.
The boundary interactions are encoded into
the boundary reflection matrix. 
Using the TBA method, we verify the flows of the conformal
BCs by computing the boundary entropies.
These flows of the BCs have direct interpretations for the 
fusion RSOS lattice models.
For super CFTs ($k=2$) we show that these flows are possible 
only for the Neveu-Schwarz sector and are consistent with the 
lattice results.
The models we considered cover a wide class of integrable models.
In particular, we show how the impurity spin is screened by electrons
for the $k$-channel Kondo model by taking $l\to\infty$ limit.
We also study the problem using an independent method based on the 
boundary roaming TBA.
Our numerical results are consistent with the boundary CFTs and RSOS 
TBA analysis.

\end{abstract} 
\rightline{EWHA-TH-9801}
\rightline{SNUTP-98-043}
\rightline{APCTP-98-012}
\end{titlepage}

\section{Introduction}

In the study of the two-dimensional quantum field
theories and statistical models, the conformal field
theories (CFT) \cite{BPZ} have provided both theoretical 
frameworks and powerful tools.
This applies to off-critical models, not to mention critical ones. 
Various off-critical quantities can be related to those of CFTs.
The two-body scattering amplitude ($S$-matrix) of a perturbed
CFT \cite{zamol} can reproduce the central 
charge and identify the perturbing field using 
the thermodynamic Bethe ansatz
(TBA) \cite{zamtba} and provide a proof that most integrable quantum 
field theories are CFTs perturbed by some relevant operator along which
the theory extends away from criticality. 
Two-dimensional lattice models with Boltzmann weights 
satisyfing the Yang-Baxter equation (YBE) also show these properties. 
For example, the order parameters such as local height probabilities (LHP)
of the restricted solid-on-solid (RSOS) models can generate critical 
exponents like the central charges of minimal CFTs \cite{ABF} and 
are related to the characters of the CFTs \cite{DJMOi}.

Such relations between critical and off-critical theories in the 
presence of boundary are main question investigated in this paper. 
There have been many interesting progresses in the quantum field theory
in the space-time with boundary.
In paricular, existence of the boundary in
$(1+1)$ dimensions changes conserved quantities, symmetries and 
integrability and requires new formulations for both CFTs and 
integrable off-critical models.
To maintain infinite dimensional conformal symmetries, the CFTs 
should introduce new quantities, the conformal boundary states, 
which have one-to-one correspondence with
conformal boundary conditions (CBC) \cite{cardyi}.

Off-critical models which can be represented by a symbolic action 
\begin{equation}
{\cal A}={\cal A}_{\tiny\rm CFT+CBC}
+\Lambda\int d^{2}x\Phi+\l\int dt \phi_{B},
\label{pbcft}
\end{equation}
can maintain the integrability if one choose the perturbing fields
carefully. 
For example, it is shown in \cite{goszam} that
the minimal CFTs perturbed by the least relevant bulk and boundary 
primary fields can be integrable.
Once the integrability is implemented, we need new quantities
called the boundary reflection amplitudes ($R$-matrices) which are the 
probability amplitudes for a particle to scatter off from the boundary.
The boundary YBE and bootstrap arguments
can determine these amplitudes completley with a given
$S$-matrix, the bulk scattering amplitude, as an input \cite{bybe,goszam}.
As the bulk TBA based on the $S$-matrix give the CFT data underlying in
the perturbed CFTs such as central charges, the $R$-matrix can be
used to find the information on the boundary conditions.
This is our main question investigated in this paper.

TBA analysis with the boundary should be a natural method for 
this purpose.
Recently, TBA for the perturbed minimal 
CFTs with certain BCs has been used to find the boundary entropies 
and corresponding CBCs \cite{LSS}.
With certain $R$-matrices, the CBCs flow into each other which are
consistent with independent study based on the roaming TBA 
\cite{zamroam}.
This result seems very interesting since it can provide a general 
method to understand the relations of the 
$R$-matrix and CFTs data in (\ref{pbcft}).
In this paper, we generalize this result on the minimal models to
general coset CFTs \cite{GKO} which include the superconformal series
and $Z_{k}$ parafermionic algebras \cite{cosetCFT}.
The minimal CFTs are only the first of many infinite series of 
these CFTs.
One can obtain a variety of off-critical integrable models 
associated with these CFTs which include many interesting ones.

These questions are not merely of academic interests. 
They can be directly related to physical quantities measurable
in experiment.
Recent progress have shown that these theoretical tools can be 
very efficient for quantitative understanding of non-perturbative
aspects of boundary problems.
For example, boundary CFTs \cite{cardyi} have been applied to 
multi-channel Kondo models \cite{afflud} 
and integrable field theories to transport
phenomena in quantum impurity problems \cite{FLS}.
The boundary entropies associated with CBCs are one of main physical 
quantities in these computations at UV and IR fixed points.
Our methods for the general coset theories with
certain limits are used to understand the flows between the 
CBCs quantitatively.
Our results will be of use to estabilish the boundary TBA method 
to compute the boundary entropy and to understand boundary 
behaviours in the intermediate boundary scale 
for a wide class of integrable models including the multi-channel
Kondo model.

This paper is orgarnized as follows. 
In sect.2, we define the general CFTs and associated lattice models. 
We derive the boundary entropies for the CBCs of these CFTs
following standard procedure. 
The boundary perturbations of these CFTs are understood in terms of 
massless scattering matrices in sect.3. These $S$-matrices are used
to compute the boundary entropies using the TBA.
In sect.4, we compute the flows of the boundary entropies using an
independent analysis, namely, boundary roaming TBA.
The boundary roaming TBA of super CFTs are obtained by analytic 
continuation of that of the supersymmetric sinh-Gordon (SShG) model.
Similar analysis is proposed for the parafermionic models.
Numerical analysis shows that the results are consistent with
those obtained from the boundary CFTs and RSOS TBA.
We conclude with some discussions in sect.5.

\section{coset CFT with boundary}

\subsection{coset CFTs}

Most rational CFTs can be expressed as coset CFTs \cite{GKO}. 
In particular, general $SU(2)$ coset CFTs ${\cal M}(k,l)$ 
represented schematically as
\[
{\cal M}(k,l)={SU(2)_k\otimes SU(2)_l\over{SU(2)_{k+l}}}
\]
with central charges 
\[
c(k,l)={3k\over{k+2}}\left[1-{2(k+2)\over{(l+2)(k+l+2)}}\right]
\]
include many important and frequently used CFTs.
Here, $SU(2)_k$ is the level $k$ Wess-Zumino-Witten (WZW) model 
with Kac-Moody algebra as an extended conformal symmetry \cite{WZW}.
Besides the minimal CFTs \cite{BPZ} ${\cal M}(1,l)$,
there are many other series with some extended symmetries
which will be of our main concern.
Our convention is that the first index $k$ of ${\cal M}(k,l)$
denotes the extended symmetries.

In particular, the superconformal theories 
${\cal M}(2,l)$ with $c=c(2,l)$ with the primary fields $\Phi_{(r,s)}$ 
$1\le r\le p-1,\ 1\le s\le p+1$ \cite{FQS}.
We will use often $p$ for $l+2$ throughout the paper.
Here, $\Phi_{(r,s)}$ is identified with $\Phi_{(p-r,p+2-s)}$.
The super CFTs and their representations can be classified into two
sectors, the Neveu-Schwarz (NS) and Ramond (R) sectors, which are
selected by antiperiodic or periodic boundary conditions on fermionic
fields, respectively. In the above notations, the primary field
$\Phi_{(r,s)}$ belongs to the (NS) or (R) sectors depending on
$r-s$ even or odd integers.

In general the coset CFTs ${\cal M}(k,l)$ ($k$ fixed and $l=1,2,\ldots$)
are extended CFTs with $Z_{k}$ parafermion currents \cite{FZ}.
The characters of the coset theories are defined as branching functions
\[
\chi^{[k]}_{t}(\tau,z)\chi^{[l]}_{r}(\tau,z)=\sum_{s=1}^{k+l+1}
B_{r,s}^{t}(\tau)\chi^{[k+l]}_{s}(\tau,z),
\]
where $\chi^{[l]}_{r}$ is the character of the highest weight
$r$ ($r=1,\ldots,l+1$) for the $SU(2)_l$ WZW model.
The primary fields for the coset CFTs, $\Phi_{r,s}^{t}$,
have three weights which take values in
\begin{eqnarray}
& &t=\vert(r-s)\ {\rm mod}\ 2k\vert+1, \quad 1\le t\le k+1,\\
\label{sector}
& &1\le r\le l+1,\quad 1\le s\le k+l+1.
\label{range}
\end{eqnarray}
The index $t$ stands for the sectors of the extended symmetries.
For example, $t=1,3$ corresponds to the (NS) and $t=2$ 
to the (R) sector of the supersymmetry ($k=2$).
Since ${\cal M}(k,l)\equiv{\cal M}(l,k)$, one can have two different
realizations for the CFTs. ${\cal M}(2,3)$ can be the third CFT
of the super CFTs or the second CFT of the $Z_{3}$ parafermion
theory by rearranging the conformal modules.  

One of the fundamental quantities in CFTs is the modular $S$-matrix
for $B_{r,s}^{t}(\tau)$ which can be obtained
by tranforming $\tau\to-1/\tau$ in the above expression \cite{DJMOii}
and using that of $SU(2)_l$ WZW model.
The results are\footnote{
Strictly speaking, the characters of the coset CFTs are linear 
combinations of the branching functions belonging to the same sector
($t=1$ and $3$ in the super CFTs). The modular $S$-matrix
for the characters will be modified except some sectors
like the (NS) where this complicacy disappears.}
\begin{equation}
{S^{[t,t']}}_{(r,s)}^{(r',s')}=
\sqrt{{8\over{(k+2)p(p+k)}}}\sin{\pi tt'\over{k+2}}
\sin{\pi rr'\over{p}} \sin{\pi ss'\over{p+k}}.
\label{scftS}
\end{equation}
The modular $S$-matrix can be simplified by choosing special 
sectors as
\begin{equation}
S_{(r,s)}^{(r',s')}=
{\rm const.}\sin{\pi rr'\over{p}} \sin{\pi ss'\over{p+k}},
\label{cosetS}
\end{equation}
where the constant factor is not of our concern since it 
will be cancelled out in the quantities of interest.

For the super CFTs, the $S$-matrices of the two sectors are 
in general complicated except the (NS)-(NS) $S$-matrix \cite{superS} 
given by
\begin{equation}
S_{(rs)}^{(r's')}={4\over{\sqrt{p(p+2)}}}\sin{\pi rr'\over{p}}
\sin{\pi ss'\over{p+2}}.
\end{equation}
For the reason explained below, we will restrict our analysis to
the (NS) sector.

\subsection{boundary conditions}

CFTs can make sense in two dimensions with a boundary 
only with well-defined CBCs classified by
Cardy \cite{cardyi}. 
With the boundary conditions on both sides of the strip $\a,\b$,
the partition function $Z_{\a\b}$ can be expressed as
\begin{equation}
Z_{\a\b}(q)=\sum_{i}n_{\a\b}^{i}\chi_{i}(\tau),
\end{equation}
where $n_{\a\b}^{i}$ denotes the number of times that the irreducible
representation $i$ occurs under the BC $\a\b$.
Using the modular transformation $\tau\to -1/\tau$,
one can reexpress this as
\[
Z_{\a\b}(q)=\sum_{j}\langle\a\vert j\rangle\langle j\vert\b\rangle
\chi_{j}(-1/\tau)
\]
from which one drives the Cardy equation,
\[
\sum_{i}S_{i}^{j}n_{\a\b}^{i}=
\langle\a\vert j\rangle\langle j\vert\b\rangle.
\]
The state $\vert\a\rangle$ satisfying this equation defines the CBC.
It is found that for each primary field $\Phi_l$ there corresponds
a CBC $\vert{\tilde l}\rangle$ which is defined in such a way that
the partition function with this CBC on one side is identified with
the character of $\Phi_l$, namely,
\begin{equation}
Z_{{\tilde 0}{\tilde h}_i}=\chi_{i}(q).
\label{bparti}
\end{equation}
This boundary state is expressed as a linear combination of the primary
states of the CFT,
\begin{equation}
\vert{\tilde h}_i\rangle=\sum_{j} {S^{j}_{i}\over{ \sqrt{S^{j}_{0}} }}
\vert j\rangle.
\end{equation}

Since $\vert 0\rangle$ defines the ground state of the CFT,
$\langle 0\vert{\tilde l}\rangle$ defines the ground degeneracy of 
the boundary state.
This boundary degeneracy $g$ is given by
\begin{equation}
g_{i}\equiv \langle 0\vert{\tilde h}_i\rangle
={S^{0}_{i}\over{\sqrt{S^{0}_{0}}}},
\label{cardy}
\end{equation}
and the boundary entropy, defined by $s_{\rm B}=\log g$,
can be completely determined by the modular $S$-matrix elements.

For the general coset CFTs, one can use Eq.(\ref{cosetS}) to get
the boundary degeneracies for a simplest sector,
\[
g_{(r,s)}^{[k,l]}={\rm const.}
{\sin{\pi r\over{l+2}}\sin{\pi s\over{k+l+2}}\over{\sqrt{
\sin{\pi\over{l+2}}\sin{\pi\over{k+l+2}}}}}.
\]
In particular, degeneracies for the boundary states 
$(1,s)$ and $(r,1)$ are given by
\begin{eqnarray}
g_{(1,s)}^{[k,l]}&=&{\rm const.}
\left({\sin{\pi\over{l+2}}\over{\sin{\pi\over{k+l+2}}}}\right)^{1/2}
\sin{\pi s\over{k+l+2}}
\label{beni}\\
g_{(r,1)}^{[k,l]}&=&{\rm const.}
\left({\sin{\pi\over{k+l+2}}\over{\sin{\pi\over{l+2}}}}\right)^{1/2}
\sin{\pi r\over{l+2}}.
\label{benii}
\end{eqnarray}
Using (\ref{scftS}) and (\ref{cardy}) one can find
the boundary degeneracies for the (NS) sector ${\tilde h}_{(1,s)}$ and 
${\tilde h}_{(r,1)}$ of the super CFTs as follows ($r,s$ odd):
\begin{eqnarray}
g_{(1,s)}^{[2,p-2]}&=&\left({16\over{p(p+2)}}\right)^{1/4}
\left({\sin{\pi\over{p}}\over{\sin{\pi\over{p+2}}}}\right)^{1/2}
\sin{\pi s\over{p+2}}\label{sbenti}\\
g_{(r,1)}^{[2,p-2]}&=&\left({16\over{p(p+2)}}\right)^{1/4}
\left({\sin{\pi\over{p+2}}\over{\sin{\pi\over{p}}}}\right)^{1/2}
\sin{\pi r\over{p}}.\label{sbentii}
\end{eqnarray}

\subsection{Fusion RSOS lattice model}

For later purposes, it is useful to have lattice model realizations
for the general coset CFTs.
The generalizations of the original RSOS model \cite{ABF}
have the Boltzmann weights $W(a,b,c,d)$ defined by four heights
at four corners of a square, each taking values in the
$A_{k+l+1}$-Dynkin diagram, where
two adjacent heights are subject to the following conditions \cite{DJMOi}
\begin{eqnarray}
& &a=1,2,\ldots,k+l+1,\nonumber\\
& &a-b=-k,-k+2,\ldots,k-2,k \label{rangei}\\
& &(a+b-k)/2=0,1,\ldots,l+1.\nonumber
\end{eqnarray}
From the $Z_2$ automorphism of the Dynkin diagram, the model is
equivalenet under the simultaneous change of
$a\to k+l+2-a$.
The LHP $P(a/b,c)$, the probability for a height to be $a$
under the boundary heights to have $(b,c)$, have been computed
and related to the branching function of ${\cal M}(k,l)$ in the
regime III, 
\begin{equation}
B_{d,a}^{e},\qquad{\rm with}\qquad e={b-c+k\over{2}}+1,\quad
d={b+c-k\over{2}}.
\label{latbran}
\end{equation}
Notice that the range of $d,e$ with Eq.(\ref{rangei}) is consistent with 
Eqs.(\ref{sector},\ref{range}). 

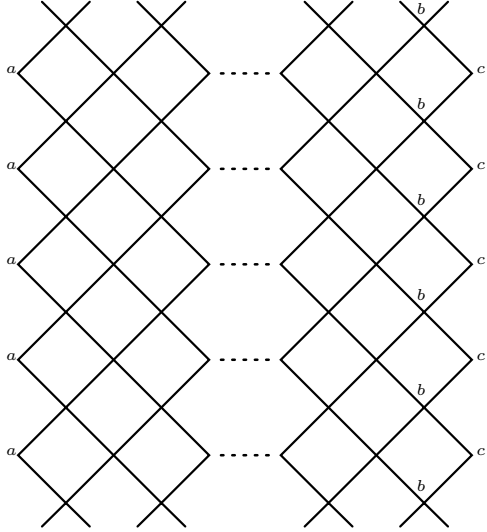
\begin{figure}[htbp]
\centering
\setlength{\unitlength}{0.0125in}
\begin{picture}(200,220)(-10,-20)
\thicklines
\path(20,200)(0,180)(20,160)
(0,140)(20,120)(0,100)(20,80)(0,60)(20,40)(0,20)(20,0)
\path(20,200)(40,180)(20,160)
(40,140)(20,120)(40,100)(20,80)(40,60)(20,40)(40,20)(20,0)
\path(60,200)(40,180)(60,160)
(40,140)(60,120)(40,100)(60,80)(40,60)(60,40)(40,20)(60,0)
\path(60,200)(80,180)(60,160)
(80,140)(60,120)(80,100)(60,80)(80,60)(60,40)(80,20)(60,0)
\path(10,210)(20,200)(30,210)
\path(50,210)(60,200)(70,210)
\path(120,210)(130,200)(140,210)
\path(160,210)(170,200)(180,210)
\path(10,-10)(20,0)(30,-10)
\path(50,-10)(60,0)(70,-10)
\path(120,-10)(130,0)(140,-10)
\path(160,-10)(170,0)(180,-10)

\dottedline{5}(85,180)(105,180)
\dottedline{5}(85,20)(105,20)
\dottedline{5}(85,140)(105,140)
\dottedline{5}(85,60)(105,60)
\dottedline{5}(85,100)(105,100)
\path(130,200)(110,180)(130,160)
(110,140)(130,120)(110,100)(130,80)(110,60)(130,40)(110,20)(130,0)
\path(130,200)(150,180)(130,160)
(150,140)(130,120)(150,100)(130,80)(150,60)(130,40)(150,20)(130,0)
\path(170,200)(150,180)(170,160)
(150,140)(170,120)(150,100)(170,80)(150,60)(170,40)(150,20)(170,0)
\path(170,200)(190,180)(170,160)
(190,140)(170,120)(190,100)(170,80)(190,60)(170,40)(190,20)(170,0)

\put(-5,180){\makebox(0,0)[lb]{\tiny$a$}}
\put(-5,140){\makebox(0,0)[lb]{\tiny$a$}}
\put(-5,100){\makebox(0,0)[lb]{\tiny$a$}}
\put(-5,60){\makebox(0,0)[lb]{\tiny$a$}}
\put(-5,20){\makebox(0,0)[lb]{\tiny$a$}}

\put(192,180){\makebox(0,0)[lb]{\tiny$c$}}
\put(192,140){\makebox(0,0)[lb]{\tiny$c$}}
\put(192,100){\makebox(0,0)[lb]{\tiny$c$}}
\put(192,60){\makebox(0,0)[lb]{\tiny$c$}}
\put(192,20){\makebox(0,0)[lb]{\tiny$c$}}

\put(167,205){\makebox(0,0)[lb]{\tiny$b$}}
\put(167,165){\makebox(0,0)[lb]{\tiny$b$}}
\put(167,125){\makebox(0,0)[lb]{\tiny$b$}}
\put(167,85){\makebox(0,0)[lb]{\tiny$b$}}
\put(167,45){\makebox(0,0)[lb]{\tiny$b$}}
\put(167,5){\makebox(0,0)[lb]{\tiny$b$}}
\end{picture}
\caption{\label{fig1}\small RSOS lattice with $(a/b,c)$ BC}
\end{figure}

It is interesting to express the boundary states in terms of the RSOS
lattice models.
It is found in \cite{salbau} that  the partition function of the RSOS
lattice model ($k=1$) with boundary heights fixed as $(a/b,c)$ 
in Fig.(\ref{fig1}) is given by
\[
Z(a/b,c)=\chi_{d,a}, 
\]
with $d$ is ${\rm inf}(b,c)$, smaller one of $b,c$.
From Eq.(\ref{bparti}), one can conclude that
$(1/b,c)$ BC of the RSOS lattice model corresponds to the 
CBC $\vert{\tilde h}_{(d,1)}\rangle$. 
Another BC we will consider is the case of free $b$ while
$a=1$ and $c$ are fixed, which corresponds to the CBC 
$\vert{\tilde h}_{(1,c)}\rangle$.

Now let us consider the boundary conditions for the general 
$k\ge 2$ cases, namely the `fusion' RSOS models.
If the argument of the conformal transformation of the strip 
to the annulus in \cite{salbau} are still valid for the fusion RSOS lattice,
the transfer matrix with fixed BC on the strip will be 
related to the corner transfer matrix.
This means that the partition function $Z(a/b,c)$ on the strip 
is related to the LHP $P(a/b,c)$, hence to the 
the branching function in Eq.(\ref{latbran}). 
Therefore, we can conjecture that analogous results 
can hold for the fusion models; namely,
$(1/b,c)$ BC of the fusion model corresponds to
the CBC $\vert{\tilde h}^{e}_{(d,1)}\rangle$ with $d,e$ given above.
Using the invariance under $c\to k+p-c$, 
we will restrict the boundary heights to $1\le b,c\le(k+p)/2$.

From (\ref{rangei}), $b$ can take three values, namely $c-2,c+2,c$ 
for $k=2$.
Plugging into (\ref{latbran}), one can find that
$b=c\pm 2$ and $b=c$ correspond to the (NS) and (R) sectors,
respectively.
One can also see that $c-a$ should be even since 
the differences of two neighboring heights should be even. 
This means the BC $(1/b,c)$ makes sense only for odd $b,c$. 
Because Eq.(\ref{sector}) with $r=d$ and $s=a$ gives $t=1,3$, 
the boundary states should always belong to the (NS) sector
with $b=c\pm 2$. 
Similarly $(1/c)$ BC with odd $c$ is identified with the CBC
$\vert{\tilde h}^{\rm\tiny NS}_{(1,c)}\rangle$.
We summarize as follows ($d,c$ are odd integers):
\begin{eqnarray}
(1/c\pm 2,c)\to \vert{\tilde h}^{\rm\tiny NS}_{(d,1)}\rangle
\quad&{\rm with}&\quad 1\le d=[(c-1)\pm 1]\le{p\over{2}}-1,
\label{bci}\\
(1/c)\to \vert{\tilde h}^{\rm\tiny NS}_{(1,c)}\rangle
\quad&{\rm with}&\quad 1\le c\le{p\over{2}}+1.
\label{bcii}
\end{eqnarray}

\section{Massless Boundary Scattering}

\subsection{Bulk RSOS TBA}

It has been claimed years ago that the minimal CFTs perturbed by the least
relevant operator are integrable and can be described by RSOS scattering
theories \cite{berlec} $S_{\tiny{\rm RSOS}(k)}(\t)$, the 
RSOS $S$-matrix whose quantum group parameter 
is given by $q=-e^{i\pi/(k+2)}$.
For the perturbed general coset CFTs, similar results have been 
obtained where the particles carry two sets of RSOS spins, namely,
$\vert K_{a,b}(\t)\rangle\otimes\vert K_{c,d}(\t)\rangle$ 
with $S$-matrices \cite{ABL}
\begin{equation}
S(\t)=S_{\tiny{\rm RSOS}(k)}(\t)\otimes S_{\tiny{\rm RSOS}(l)}(\t).
\label{smat}
\end{equation}
The first set of RSOS spins (${a,b}$) acting on the first $S$-matrix 
is considered as the index for internal symmetries such as
supersymmetry ($k=2$).
These particles, `kinks', are obtained by restricting
multi-soliton Hilbert space when the quantum group
parameter $q$ is a root of unity.
These massive theories correspond to the perturbed CFTs with negative
coefficients. We denote this by ${\cal M}A^{(-)}(k,l)$.

If the coefficients of the perturbing operator are positive, 
the perturbed CFTs will flow between two fixed points \cite{rgflow}.
These flows of the central charges have been reproduced using the 
above $S$-matrix in the thermodynamic
Bethe ansatz (TBA) analysis by changing only
the dispersion relation to 
$E=\pm P=\pm Me^{\pm\t}$ \cite{zamflow}
which means the left-moving ($-$) and right-moving ($+$)
massless particles.
$M$ is a mass scale which is connected with the dimensionful
perturbing parameter $\Lambda$.
These theories, denoted by ${\cal M}A^{(+)}(k,l)$, with
$S_{LL}$ and $S_{RR}$ given by Eq.(\ref{smat}) and with 
appropriate $S_{LR}$ are interpolating
two adjacent CFT series in the following way \cite{zamcoset}:
\begin{equation}
{\cal M}(k,l)\quad\to\quad {\cal M}(k,l-k).
\label{Bflow}
\end{equation}
Notice that there are more than one sequences of the flows within
CFTs with fixed $k\ge 2$.

Furthermore, it has been claimed in \cite{zamcft}
that in the vanishing limit of $\Lambda$, 
one can still preserve the massless kink spectrum
along with the RSOS $S$-matrices. Since the perturbed CFTs in the
limit of vanishing perturbations are obviously the CFTs, these scattering
theories can describe the CFTs.
Only difference from the ${\cal M}A^{(+)}(k,l)$ is that $S_{LR}=1$, i.e.
trivial scattering between $L$ and $R$-movers.
All these theories are invariant under $k\leftrightarrow l$.

Non-pertubative results can be obtained by the TBA.
It is very complicated to derive the TBA equations for the RSOS 
$S$-matrices and is not of our concern. 
For detailed derivations, see \cite{bazres}.
Instead, we sketch briefly the conceptual aspects only which
will be useful to understand the boundary cases.
With nondiagonal $S$-matrices, one needs to diagonalize
the transfer matrices arising in the periodic BC.
The eigenvalues depends on the particle rapidities as well as 
the `magnonic' mode which, in turn, satisfies some constraint equation.
To define this constraint, one needs to introduce another mode and
so on. For RSOS($k$), one needs $k-1$ massless magnonic modes.
By interpreting these modes as massless particles and the constraints
as the periodic BC, one can transform the nondiagonal problem into the
that of diagonal scattering theories. The rest of the derivation is
straightforwardly standard.
Since only the first magnon rapidity will enter to define the 
eigenvalues, the massive particle scatters with the first magnon and
the first with the second etc.

For the general cases with $S$-matrix (\ref{smat}), 
the transfer matrix will be also the tensor product form and the 
eigenvalues are products of two factors which have two sets of magnons
($k-1$ for RSOS($k$) and $l-1$ RSOS($l$)).
In the effective diagonal TBA, the massive particle scatters
with two first magnons and 
the first with the second for a given factor and so on. 
This is represented in the TBA diagram of Fig.\ref{fig2}(a)
TBA for ${\cal M}A^{(+)}(k,l)$ and ${\cal M}(k,l)$ are
conjectured similarly and represented in Fig.\ref{fig2}(b) and (c).
Here, the index $k$ is the smaller of $k,l$.
The exchange of $k$ and $l$ does not change the TBA.

\begin{figure}
\centering
\setlength{\unitlength}{0.0125in}
\begin{picture}(350,50)(-25,0)
\put(-25,5.5){\small (a)}
\put(0,6.5){$\bigcirc$}
\put(12,10){\line(1,0){39}}
\put(50,6.5){$\bigcirc$}
\dottedline{4}(70,10)(90,10)
\put(100,6.5){$\bigotimes$}
\put(111,10){\line(1,0){40}}
\put(150,6.5){$\bigcirc$}
\dottedline{4}(170,10)(190,10)
\put(200,6.5){$\bigcirc$}
\put(212,10){\line(1,0){39}}
\put(250,6.5){$\bigcirc$}
\dottedline{4}(270,10)(290,10)
\put(300,6.5){$\bigcirc$}
\put(4,-2){\makebox(0,0)[lb]{\tiny $1$}}
\put(54,-2){\makebox(0,0)[lb]{\tiny $2$}}
\put(104,-2){\makebox(0,0)[lb]{\tiny $k$}}
\put(150,-2){\makebox(0,0)[lb]{\tiny $k+1$}}
\put(204,-2){\makebox(0,0)[lb]{\tiny $l$}}
\put(250,-2){\makebox(0,0)[lb]{\tiny $l+1$}}
\put(295,-2){\makebox(0,0)[lb]{\tiny $k+l-1$}}
\put(95,20){\makebox(0,0)[lb]{\tiny $r\cosh\t$}}
\end{picture}
\begin{picture}(350,50)(-25,0)
\put(-25,5.5){\small (b)}
\put(0,6.5){$\bigcirc$}
\put(12,10){\line(1,0){39}}
\put(50,6.5){$\bigcirc$}
\dottedline{4}(70,10)(90,10)
\put(100,6.5){$\bigotimes$}
\put(111,10){\line(1,0){40}}
\put(150,6.5){$\bigcirc$}
\dottedline{4}(170,10)(190,10)
\put(200,6.5){$\bigotimes$}
\put(211,10){\line(1,0){40}}
\put(250,6.5){$\bigcirc$}
\dottedline{4}(270,10)(290,10)
\put(300,6.5){$\bigcirc$}
\put(4,-2){\makebox(0,0)[lb]{\tiny $1$}}
\put(54,-2){\makebox(0,0)[lb]{\tiny $2$}}
\put(104,-2){\makebox(0,0)[lb]{\tiny $k$}}
\put(150,-2){\makebox(0,0)[lb]{\tiny $k+1$}}
\put(204,-2){\makebox(0,0)[lb]{\tiny $l$}}
\put(250,-2){\makebox(0,0)[lb]{\tiny $l+1$}}
\put(295,-2){\makebox(0,0)[lb]{\tiny $k+l-1$}}
\put(100,25){\makebox(0,0)[lb]{\tiny $\half re^{\t}$}}
\put(200,25){\makebox(0,0)[lb]{\tiny $\half re^{-\t}$}}
\end{picture}
\begin{picture}(350,50)(-25,0)
\put(-25,5.5){\small (c)}
\put(0,6.5){$\bigcirc$}
\put(12,10){\line(1,0){39}}
\put(50,6.5){$\bigcirc$}
\dottedline{4}(70,10)(90,10)
\put(100,6.5){$\bigotimes$}
\put(111,10){\line(1,0){40}}
\put(150,6.5){$\bigcirc$}
\dottedline{4}(170,10)(190,10)
\put(200,6.5){$\bigcirc$}
\put(212,10){\line(1,0){39}}
\put(250,6.5){$\bigcirc$}
\dottedline{4}(270,10)(290,10)
\put(300,6.5){$\bigcirc$}
\put(4,-2){\makebox(0,0)[lb]{\tiny $1$}}
\put(54,-2){\makebox(0,0)[lb]{\tiny $2$}}
\put(104,-2){\makebox(0,0)[lb]{\tiny $k$}}
\put(150,-2){\makebox(0,0)[lb]{\tiny $k+1$}}
\put(204,-2){\makebox(0,0)[lb]{\tiny $l$}}
\put(250,-2){\makebox(0,0)[lb]{\tiny $l+1$}}
\put(295,-2){\makebox(0,0)[lb]{\tiny $k+l-1$}}
\put(100,25){\makebox(0,0)[lb]{\tiny $\half re^{\t}$}}
\end{picture}
\caption{\label{fig2}\small TBA diagrams
(a) ${\cal M}A^{(-)}(k,l)$ (b) ${\cal M}A^{(+)}(k,l)$ 
(c) ${\cal M}(k,l)$}
\end{figure}

Explicit TBA equations are expressed as follows:
\[
\ep_{a}(\t)=\nu_{a}(\t)-\sum_{b=1}^{k+l-1} l_{ab}\varphi*L_{b}(\t),
\quad a=1,\ldots,k+l-1,
\]
where
\begin{eqnarray*}
\varphi(\t)&=&{1\over{\cosh\t}}\\
L_{b}(\t)&=&\log\left(1+e^{-\ep_{b}(\t)}\right)\\
f*g(\t)&=&\int{d\t'\over{2\pi}}f(\t-\t')g(\t'),
\end{eqnarray*}
and the source terms $\nu_{a}(\t)$ are given by
\begin{eqnarray}
\nu_{a}(\t)&=&\de_{ak}r\cosh\t\quad{\rm for}\quad {\cal M}A^{(-)}(k,l)
\nonumber\\
\nu_{a}(\t)&=&\de_{ak}r{e^{\t}\over{2}}+\de_{al}r{e^{-\t}\over{2}}
\quad{\rm for}\quad {\cal M}A^{(+)}(k,l)
\label{Brgtba}\\
\nu_{a}(\t)&=&\de_{ak}r{e^{\t}\over{2}}
\quad{\rm for}\quad {\cal M}(k,l)
\end{eqnarray}
where a dimensionless parameter $r$ definded by $M/T$ with temperature
$T$ interpolates the UV ($r\to 0$) and IR ($r\to\infty$) limits.
$l_{ab}$ is the incidence matrix whose elements are $1$ if
two nodes $a,b$ are connected in the TBA diagram
Fig.(\ref{fig2}) or $0$ otherwise.

In the UV and IR limits, $\ep_{a}$'s only at $\t=\pm\infty,0$ are 
important and can be determined by simple algebraic equations.
Two of the massless TBA have the same solutions in the UV limit
\cite{bazres,zamcoset}
\begin{eqnarray}
1+e^{-\ep_a(-\infty)}&=&\left[{\sin{\pi(a+1)\over{k+l+2}}\over{
\sin{\pi\over{k+l+2}}}}\right]^2,\quad 1\le a\le k+l-1
\label{soli}\\
1+e^{-\ep_a(\infty)}&=&\left[{\sin{\pi(a+1-k)\over{l+2}}\over{
\sin{\pi\over{l+2}}}}\right]^2,\quad k\le a\le k+l-1
\label{solii}\\
1+e^{-\ep_a(\infty)}&=&\left[{\sin{\pi(a+1)\over{k+2}}\over{
\sin{\pi\over{k+2}}}}\right]^2,\quad 1\le a\le k,
\label{soliii}
\end{eqnarray}
while the IR behaviours are all different. 
${\cal M}A^{(-)}(k,l)$ becomes massive with $e^{-\ep_{a}}=0$.
${\cal M}(k,l)$ remains same and independent of $r$
while ${\cal M}A^{(+)}(k,l)$ generates the flows (\ref{Bflow})
where the $\ep_{a}$'s are given as above with replacing $l\to l-k$.  

\subsection{Boundary RSOS TBA}

Now we introduce the boundary. The formal action is in the form of
Eq.(\ref{pbcft}).
The perturbed CFTs can be well-defined only after specifying the CBCs.
Once the integrability is maintained by specific BCs,
the boundary $R$-matrix, obtained by the boundary YBE,
can be used in the boundary TBA to compute the entropies \cite{btba}.
For diagonal $S$- and $R$-matrices,
\begin{eqnarray*}
\log\langle B_{\a}\vert 0\rangle\langle 0\vert B_{\b}\rangle&=&
\int_{-\infty}^{\infty}{d\t\over{2\pi}}\kappa_{\a\b}(\t)
\log\left(1+e^{-\ep(\t)}\right)+{\rm const.}\\
\ep(\t)&=&\half re^{\t}+\phi*L(\t)\\
\kappa_{\a\b}(\t)&=&{1\over{i}}{d\over{d\t}}
\log\left[R_{\a}(\t-\t_{B_{\a}})R_{\b}(\t-\t_{B_{\b}})\right],
\end{eqnarray*}
where `boundary rapidity' $\t_{\tiny\rm B}$ is defined by
$m_{B}=Me^{\t_{\tiny\rm B}}$ where the boundary mass scale $m_B$ is
a certain power of $\l$ in (\ref{pbcft}).
The UV (IR) limit is $\t_B\to-\infty(\infty)$.

The simplest example is ${\cal M}(1,1)$, namely the Ising model 
with $R$ given by
\[
R(\t-\t_B)=-i\tanh\left[\half(\t-\t_B)-{i\pi\over{4}}\right]
\]
which produces the boundary entropy 
$s_{\tiny\rm B}=\log\langle 0\vert B\rangle$ as
\[
s_{\tiny\rm B}=\int_{-\infty}^{\infty}{d\t\over{2\pi}}
{\log\left(1+e^{-\ep(\t)}\right)\over{\cosh(\t-\t_B)}}+{\rm const.}
\]
This gives the correct flow of the boundary degeneracy,
$g_{\tiny\rm UV}/g_{\tiny\rm IR}=\sqrt{2}$.

Generalizing this result to the RSOS($k\ge 2$) theory is nontrivial. 
The authors in \cite{LSS} claimed that
the boundary perturbation of the minimal CFT with the CBC
$\vert{\tilde h}_{(1,a)}\rangle$ is the quantum group reduction of
the massless limit of the boundary sine-Gordon model with
the anisotropic spin-$j$ Kondo interaction at boundary. 
Therefore, the boundary $R$-matrices of the boundary RSOS($k$) theory 
are given by the RSOS version of the $R$-matrix of the Kondo model
given in \cite{fen}:
\begin{eqnarray}
R(\t)&=&1,\qquad\qquad\qquad\qquad a=1\nonumber\\
R(\t)&=&-i\tanh\left({\t-\t_{\tiny\rm B}\over{2}}-{i\pi\over{4}}\right),
\quad a=2\nonumber\\
R(\t)&=&R^{1/2,(a-2)/2}(\t-\t_{\tiny\rm B}),\quad
3\le a\le{p+1\over{2}},
\label{Rkondo}
\end{eqnarray}
with $a=2j+1$.
Notice $j=1/2$ ($a=2$) where the boundary spin is fixed at spin-$1/2$ 
gives the same $R$ as the Ising model.
Based on these $R$-matrices,
the boundary entropy for the CBC $\vert{\tilde h}_{(1,a)}\rangle$ 
has been conjectured as 
\begin{equation}
s_{\tiny\rm B}^{(a)}=\int{d\t\over{2\pi}}
{\log\left(1+e^{-\ep_{a-1}(\t)}\right)
\over{ \cosh(\t-\t_{\tiny\rm B})}},
\label{bentropy}
\end{equation}
where $\ep_{a}$'s are determined by usual RSOS bulk TBA.
With solutions of the bulk TBA,
the ratios of the boundary degeneracies become
\begin{equation}
{g_{\tiny\rm UV}^{(a)}\over{g_{\tiny\rm IR}^{(a)}}}=
{1+e^{-\ep_{a-1}(-\infty)}\over{1+e^{-\ep_{a-1}(\infty)} }}.
\label{uiratio}
\end{equation}

These results can be a guideline for the general coset CFTs. 
Considering the bulk $S$-matrix (\ref{smat}),
one can look for the $R$-matrix in the form of
\[
R(\t)=R_{\tiny{\rm RSOS}(k)}(\t)\otimes R_{\tiny{\rm RSOS}(l)}(\t).
\]
The first factor is related to the internal symmetry.
The fractional supersymmetry is defined in the $S$-matrix of 
the RSOS($k$) \cite{berlec}.
In the bulk conformal limit $\Lambda\to 0$, this symmetry will
remain as extended conformal symmetries generated by the $Z_k$
parafermion.  
Even in the presence of boundary, we require that the CFTs 
should preserve these extended symmetries along with the conformal ones.
Among many solutions \cite{ahn} for the boundary $R$-matrix of 
the RSOS($k$) model, no one can preserve the conserved 
currrents of fractional spin except the trivial case of $R\propto{\bf 1}$.
This is consistent with the results in \cite{warner} 
where the $R$-matrix of massless particles with topological charges  
corresponding to the Dirichlet BC is ${\bf 1}$. 
We will use the $R$-matrix (\ref{Rkondo}) for the second $R$.
To summarize, our conjecture for the $R$-matrix of the CBC 
$\vert{\tilde h}_{(1,a)}\rangle$ is ${\bf 1}\otimes R^{1/2,(a-2)/2}$.

The boundary TBA can be derived similarly. 
While Boundary part will be the same as before, 
Eqs.(\ref{bentropy},\ref{uiratio}), since the reflection amplitudes
do not change, bulk part will be described by TBA of ${\cal M}(k,l)$.  
Let's consider for the case of $a>k$ first.
Plugging Eqs.(\ref{soli},\ref{solii}) into (\ref{uiratio}), 
one gets
\begin{equation}
{g_{\tiny\rm UV}^{(a)}\over{g_{\tiny\rm IR}^{(a)}}}
={\sin{\pi a\over{k+l+2}}\over{\sin{\pi\over{k+l+2}}}}
{\sin{\pi\over{l+2}}\over{\sin{\pi(a-k)\over{l+2}}}}.
\label{resulti}
\end{equation}
Comparing (\ref{resulti}) with Eqs.(\ref{beni},\ref{benii}), 
we find that the boundary reflection matrices generate the flows
\begin{equation}
\vert{\tilde h}_{(1,a)}\rangle\to
\vert{\tilde h}_{(a-k,1)}\rangle.
\label{bflowii}
\end{equation}
This means the UV CBC $\vert{\tilde h}_{(1,a)}\rangle$ 
of the perturbed boundary CFTs ${\cal M}(k,l)$ 
changes to $\vert{\tilde h}_{(a-k,1)}\rangle$ in the IR ($a\ge k+1$).

It will be more instructive to intepret this result in the 
fusion RSOS lattice model language. 
For $k+1\le a\le (k+l+2)/2$, Eq.(\ref{bflowii}) means simply 
the flow from $(1/a)$ BC to $(1/a-k,a)$.
For $(k+l+2)/2\le a\le k+l+1$ the CBC $\vert{\tilde h}_{(1,a)}\rangle$ 
and $\vert{\tilde h}_{(a-k,1)}\rangle$ correspond to 
the lattice BCs $(1/a')$ and $(1/a'+k,a')$ with $a'=k+l+2-a$ due to
the $Z_2$ symmetry.  According to Eq.(\ref{latbran}), these 
boundary states are identified with the CBC 
$\vert{\tilde h}_{(1,a')}\rangle$ and $\vert{\tilde h}_{(a',1)}\rangle$,
respectively.  
Therefore, the boundary TBA results give two flows
with $k+1\le a\le (k+l+2)/2$, 
\begin{eqnarray}
\vert{\tilde h}_{(1,a)}\rangle\to
\vert{\tilde h}_{(a-k,1)}\rangle\quad&{\rm or}&\quad
(1/a)\to(1/a-k,a) \label{sumi}\\
\vert{\tilde h}_{(1,a)}\rangle\to
\vert{\tilde h}_{(a,1)}\rangle\qquad&{\rm or}&\quad
(1/a)\to(1/a+k,a).\label{sumii}
\end{eqnarray}
Our results reproduce \cite{LSS} for $k=1$ where
it is claimed that these flows are associated with
the signs of $\l$ in (\ref{pbcft}).
Since $\l$ is only free parameter, this conclusion should be
also true for general cases.

For the case of $a\le k$, 
the boundary degeneracies can be
computed from (\ref{soli},\ref{soliii}). 
The result is
\begin{equation}
{g_{\tiny\rm UV}^{(a)}\over{g_{\tiny\rm IR}^{(a)}}}
={\sin{\pi a\over{k+l+2}}\over{\sin{\pi\over{k+l+2}}}}
{\sin{\pi\over{k+2}}\over{\sin{\pi a\over{k+2}}}}
={g^{[l,k]}_{(1,a)}\over{g^{[l,k]}_{(a,1)}}},
\label{resultii}
\end{equation}
where we used Eqs.(\ref{beni},\ref{benii}) at the last equality.
Notice that the indices $k$ and $l$ are switched.
This means the the BC flow for the $Z_{l}$, not $Z_k$, parafermion CFTs; 
\[
\vert{\tilde h}_{(1,a)}\rangle\to\vert{\tilde h}_{(a,1)}\rangle.
\]
For the lattice model, this is the flow $(1/a)\to(1/a+k,a)$
and with (\ref{sumii}) $a$ is now extended to $1\le a\le (k+l+2)/2$.

Our model covers a wide range of integrable models for each $k$ and $l$.
In particular, the result for $l\to\infty$ limit is interesting 
and has also physical applicability.
The coset CFT in this limit becomes the level-$k$ WZW model 
with boundary interaction. 
This is the multi-channel Kondo model of the spin
current considered in \cite{afflud}.
For $a=2j+1>k$, Eq.(\ref{resulti}) gives
\[
{g_{\tiny\rm UV}^{(a)}\over{g_{\tiny\rm IR}^{(a)}}}
={a\over{a-k}},\quad a>k,
\]
which means that the spin degeneracy of $2j+1$ at the UV limit
flows into $2(j-k/2)+1$ in the IR.
This is nothing but the screening effect of the impurity spin 
by $k$-channel electrons in the underscreend case which can 
be obtained by `fusion hypothesis' \cite{afflud}.
The case of $2j+1\le k$ is more interesting. 
Eq.(\ref{resultii}) gives
\[
{g_{\tiny\rm UV}^{(a)}\over{g_{\tiny\rm IR}^{(a)}}}
=a\cdot {\sin{\pi\over{k+2}}\over{\sin{\pi a\over{k+2}}}}.
\]
This is exactly the result for the overscreened Kondo model previously
obtained by the Bethe ansatz and boundary CFT results.

Our TBA result can be used to understand the flows quantitatively.
We show the numerical result of the boundary entropies 
for several boundary spins ($k=3$ and $l=4$) as the boundary
scale changes in Fig.3.
This graph illustrates nonperturbatively
the `$g$-theorem' that the boundary entropy 
always decreases as the system goes from UV to IR \cite{afflud}.

\subsection{Massless flows in the Bulk and Boundary}
We considered so far the case of $\Lambda=0,\l\neq 0$ which
shows only the boundary flows. 
Natural extension will be the simultaneous flow of the bulk and 
boundary. 
Due to nontrivial scattering between the right- and left-movers,
the bulk TBA is given by (\ref{Brgtba}). 
For the boundary entropy, an educated guess is
\begin{equation}
s_{\tiny\rm B}^{(a)}=\half\int_{-\infty}^{\infty}{d\t\over{2\pi}}\left[
{\log\left(1+e^{-\ep_{a-1}(\t)}\right)\over{\cosh(\t-\t_{\tiny\rm B})}}
+{\log\left(1+e^{-\ep_{k+l+1-a}(\t)}\right)
\over{\cosh(\t+\t_{\tiny\rm B})}} +{\rm const.}\right].
\label{BBent}
\end{equation}
This is the same conjecture used in \cite{LSS}.

Analysis of this boundary TBA is a little more complicated.
In the bulk UV limit $r=M/T\to 0$, 
we define $\t_{\tiny\rm B}={\overline\t}+\b_{\tiny\rm B}$ 
with ${\overline\t}\to\infty$ so that $re^{\overline\t}$ is finite.
Then redefine the rapidity as $\t=\pm({\overline\t}+\b)$
for $R (L)$-movers.
For $R$-movers, the source term at node $l$ vanishes
reducing $\nu_{a}$'s to those of ${\cal M}(k,l)$
and the second term in (\ref{BBent}) also vanishes and the boundary 
entropy is the half of ${\cal M}(k,l)$.
For $L$-movers, the source term at node $k$ 
and the first term in (\ref{BBent}) vanish. 
The resulting TBA and boundary entropy are the same as those by
$R$-movers using the obvious symmetry of the TBA diagram
and the solutions (\ref{soli},\ref{solii}) under $k\leftrightarrow l$.
These two contributions cancel the half in front of the entropy
and gives the same formula as before.
This result is expected since the two TBA systems are equivalent at $UV$.
 
Now consider the opposite limit, i.e. the bulk IR limit $M/T\to\infty$. 
Since $r\to\infty$, we redefine the rapidities differently, namely,
$\t_{\tiny\rm B}=-{\overline\t}+\b_{\tiny\rm B}$
and $\t=\pm({\overline\t}-\b)$ with ${\overline\t}\to\infty$. 
For $R$-movers, the source term at node $k$ becomes infinite
and the TBA diagram for ${\cal M}A^{(+)}(k,l)$ is cut at $k$.
The remaining diagram is that of ${\cal M}(k,l-k)$
and the first term in (\ref{BBent}) vanishes.
With similar result for the $L$-movers,
one gets the boundary TBA system with $l\to l-k$.
Therefore, we get the boundary flows for the IR CFT ${\cal M}A^{(+)}(k,l)$.

Fixing the boundary scale $m_B$ and varying the bulk scale shows another
interesting behaviour. For example, we consider (\ref{pbcft}) with
CBC $\vert{\tilde h}_{(1,a)}\rangle$ and $\l=0$.
In the bulk UV limit, 
the boundary degeneracy is given as before by
\[
g_{\tiny\rm UV}^{(a)}
={\rm const.}\left({\sin{\pi a\over{k+l+2}}\over{\sin{\pi\over{k+l+2}}}}
\right).
\]
In the bulk IR limit with $r\to\infty$, the source terms diverge
and the $k$ nodes from each end of the TBA diagram should be
removed. The solutions can be obtained from (\ref{soli}) with
replacing $a\to a-k$ and $k+l+2\to l-k+2$ and the boundary entropy
becomes
\[
g_{\tiny\rm IR}^{(a)}
={\rm const.}\left({\sin{\pi(a-k)\over{l-k+2}}\over{\sin{\pi\over{l-k+2}}}}
\right).
\]
Using Eqs.(\ref{beni},\ref{benii}), one can confirm that these are
the flows
\[
\vert{\tilde h}_{(1,a)}^{[k,l]}\rangle\to
\vert{\tilde h}_{(a-k,1)}^{[k,l-k]}\rangle.
\]

\section{Roaming on Boundary}

\subsection{super Roaming TBA}

Roaming model \cite{zamroam} is obtained by taking analytic continuation
of the coupling constant of the sinh-Gordon model which generates
all the minimal CFTs in one equation.
As suggested in \cite{ADM}, this model can be used to describe 
the correlation functions via form factors of all the minimal series
and their perturbations.
Its application to the boundary problem is also tested in \cite{LSS}.
The supersymmetric version of this model is a nondiagonal theory
and its bulk TBA has been derived using `the free fermion condition' in
\cite{ahni}. 

The boundary reflection amplitude of the SShG model has been studied 
in \cite{ahnkoo}.
Its boundary TBA for the $R$-matrix preserving the supersymmetry 
is found \cite{ahnii} as
\begin{eqnarray*}
\ep_{1}(\t)&=&r\cosh\t-\P*L_{2}(\t)\\
\ep_{2}(\t)&=&-\P*L_{1}(\t),
\end{eqnarray*}
with the kernel 
\[
\P(\t)={4\cosh\t\sin(\pi\a)\over{\cosh 2\t-\cos(2\pi\a)}}.
\]
The boundary entropy is given by
\begin{equation}
s_{B}={1\over{4\pi}}\int d\t\kappa(\t)\log\left(1+e^{-\ep_{1}(\t)}
\right),
\label{sbent}
\end{equation}
where $\kappa(\t)=-i\partial_{\t}\log R$.
By suppressing unnecessary parameters, $\kappa(\t)$ is given by
\[
\kappa(\t)={2\cosh f\cosh\t\over{\cosh^2\t+\sinh^2 f}},
\]
where $f$ is a dimensionless parameter determined by the mass scale
of the bulk theory and the dimensionful parameter in the boundary
potential.
This is the same as the one appears in the roaming for the minimal model.

By taking an analytic continuation of the SShG coupling constant,
\[
\pi\a={\pi\over{2}}\pm\t_{0}i\qquad{\rm with}\qquad
\t_{0}>>1,
\]
we obtain the roaming TBA
with new kernel 
\[
\P(\t)={1\over{\cosh(\t-\t_0)}}+{1\over{\cosh(\t+\t_0)}}. 
\]
Our derivation of the bulk roaming TBA is exactly the one 
conjectured in \cite{martins}
and generates the superconformal series with Witten index $1$.
The boundary entropies for super CFTs can be obtained numerically
using (\ref{sbent}) which are plotted in Fig.4 for 
$p=4,6,8$
Analyzing this figure, we can confirm that these values at the 
plateaus are consistent with $s_B=\log g$ with $g$ given in 
Eqs.(\ref{sbenti},\ref{sbentii}).
We can find that there are two types of boundary 
flows, all in the (NS) sector ($p=4n$ or $4n+2$):
\begin{eqnarray*}
\vert {\tilde h}_{1,s}\rangle &\to& \vert {\tilde h}_{s-2,1} \rangle
\quad{\rm for}\quad s=3,5,\ldots,2n+1,\\ 
\vert {\tilde h}_{1,s}\rangle &\to& \vert {\tilde h}_{s,1} \rangle
\quad{\rm for}\quad s=3,5,\ldots,2n+1. 
\end{eqnarray*}
These are consistent with Eqs.(\ref{bci},\ref{bcii}) based on the boundary
CFTs and with Eqs.(\ref{sumi},\ref{sumii}) based on the RSOS
TBA analysis.

\subsection{boundary roaming for general coset}

Bulk roaming TBA equations for the general coset CFTs
are conjectured in \cite{dorrav} as follows 
(We will consider only the $SU(2)$ coset CFTs.): 
\begin{eqnarray*}
\ep^{(i)}(\t)&=&\nu^{(i)}(\t)-\varphi_{-}*L^{(i-1)}(\t)
-\varphi_{+}*L^{(i+1)}(\t),\\
\nu^{(i)}(\t)&=&\half (\de_{i,0}r e^{-\t}+\de_{i,s}r e^{\t}),
\quad i=0,\ldots,k-1,
\end{eqnarray*}
where the index is defined cyclic $i\equiv i+k$ and
$\varphi_{\pm}(\t)=1/\cosh(\t\pm\t_0)$ and $s=0,\ldots,k-1$.
Without knowing any $S$-matrix interpretation of these equations,
we can rely on our previous experience to conjecture the 
boundary roaming TBAs. 
Our conjecture is that the bulk part is the same as before
and the boundary entropy is given by Eq.(\ref{sbent}) 
with $\ep_{1}$ replaced by $\ep^{(0)}$.
The bulk TBA is claimed to generate the roaming 
\[
\ldots\to{\cal M}(k,2k+s)\to{\cal M}(k,k+s)\to{\cal M}(k,s).
\]

Our interest is the boundary entropy generated by the general
roaming TBA.
For example, we study numerically the boundary entropies 
for $k=3$ and $s=0$ and plot the result in Fig.5. 
From this, we conclude the following BC flows:
\begin{eqnarray*}
\vert {\tilde h}_{1,s}\rangle &\to& \vert {\tilde h}_{s-3,1} \rangle
\quad{\rm for}\quad s=4,7,\ldots,3n+1,\\
\vert {\tilde h}_{1,s}\rangle &\to& \vert {\tilde h}_{s,1} \rangle
\quad{\rm for}\quad s=4,7,\ldots,3n+1.
\end{eqnarray*}
Again, this result is consistent with (\ref{sumi},\ref{sumii})
based on the boundary CFTs and RSOS scattering theories.

In Fig.4 and Fig.5, the second type flows $(1,s)\to(s,1)$ are
`inverted' flows in the sense that 
the boundary entropies increase as the scale decreases.
While this is not forbidden by the $g$-theorem since we are dealing with
theories with complex coupling constant, hence complex dimensions,
the origin is mysterious considering the roaming flows satisfy
the `$c$-theorem' faithfully.

\section{Conclusion}

In this paper, we have investigated a wide class of massless scattering
theories originated as perturbed coset CFTs.
The boundary scattering amplitudes are used to find the boundary 
entropies through the boundary TBA methods.
Our key result is to verify the flows of the BCs both for the boundary
CFTs and for the fusion lattice models.
These results are cross-checked with the extended roaming TBAs.

Several points are not clarified.
First of all, we have considered only the (NS) sector of
the super CFTs and the corresponding results make sense only
in that sector. For the fusion lattice model, it is shown that
this is only possible case.
The lattice realization of the (R) sector and the analysis
based on the boundary CFT and TBA remain to be resolved.
Even for the (NS), we have considered only special BCs, namely
$(r,1)$ and $(1,s)$, equivalently $(1/a)$ and $(1/b,c)$ 
for the lattice model.
The $R$-matrix we used describes the flows within this subset 
of BCs. 
Similarly, for $k>2$, we considered only special sector and special
type of the BCs.
It would be very interesting to find more general reflection matrices
which can generate flows between other BCs.

We considered a wide class of integrable models, `an integrable zoo'
\cite{ABL}.
Putting a boundary for the zoo is a quite interesting project.
In particular, various interesting results on the boundary 
can be obtained by taking various limits.
The cases of $k=4,\infty$ with $l\to\infty$ give the bulk CFT with
$c=2,3$, which can be realized with two and three free bosons.
The boundary behaviour of these theories may be interesting 
for string theory formulation.

Good agreement between the two results based on the RSOS and 
roaming TBAs suggest that the roaming limit of the 
super sinh-Gordon model with simple $S$-matrix can be useful to compute 
off-shell quantities such as correlators
for more physically relevant models, such as multi-channel Kondo models.

\section*{Acknowledgment}
We thank F. Lesage for explaining his work and
CA thanks A. Fring for interesting conversations and Alexander
von Humbodt foundation for financial support.
This work is supported in part by  KOSEF 961-0201-006-2 and BSRI 97-2427 
(CA) and BSRI 97-2434 (CR) and by a grant from KOSEF through CTP/SNU.
We also thank APCTP for sponsoring our program.
\vskip 1cm
\leftline{\Large\bf Figure Caption}
\begin{itemize}
\item Fig.3: Flows of the boundary entropy as boundary scale changes
for $k=3,l=4$ ($s_B$ vs. $\t_{B}$) for various boundary states $a$.

\item Fig.4: Roaming of the boundary entropies for
$k=2,s=0$ with $\t_0=40$ ($f$ vs. $s_B$).
\begin{enumerate}
\item $p=4$:
\begin{enumerate}
\item $\vert {\tilde h}_{1,3}\rangle$ to $\vert {\tilde h}_{1,1} \rangle$
($0.693..\to 0.$)
\end{enumerate}
\item $p=6$:
\begin{enumerate}
\item  $\vert {\tilde h}_{1,3}\rangle$ to $\vert {\tilde h}_{1,1} \rangle$
($0.881..\to 0.$)
\item $\vert {\tilde h}_{1,3}\rangle$ to $\vert {\tilde h}_{3,1}\rangle$
($0.881.. \to 0.693..$)
\end{enumerate}
\item $p=8$:
\begin{enumerate}
\item  $\vert {\tilde h}_{1,5}\rangle$ to $\vert {\tilde h}_{3,1} \rangle$
($1.173..\to 0.881..$),
$\vert {\tilde h}_{1,3}\rangle$ to $\vert {\tilde h}_{1,1} \rangle$
($.962..\to 0.$)
\item $\vert {\tilde h}_{1,3}\rangle$ to $\vert {\tilde h}_{3,1}\rangle$
($.962..\to 0.881..$)
\end{enumerate}
\end{enumerate}

\item Fig.5: Roaming of boundary entropies for $k=3,s=0$ 
with $\t_0=50$ ($f$ vs. $s_{\rm\tiny B}$).
\begin{enumerate}
\item $p=5$:
\begin{enumerate}
\item $\vert {\tilde h}_{1,4}\rangle$ to $\vert {\tilde h}_{1,1} \rangle$
($0.9605..\to 0.$)
\end{enumerate}
\item $p=8$:
\begin{enumerate}
\item  $\vert {\tilde h}_{1,4}\rangle$ to $\vert {\tilde h}_{1,1} \rangle$
($1.172..\to 0.$)
\item $\vert {\tilde h}_{1,4}\rangle$ to $\vert {\tilde h}_{4,1}\rangle$
($1.172.. \to 0.96054..$)
\end{enumerate}
\item $p=11$:
\begin{enumerate}
\item  $\vert {\tilde h}_{1,7}\rangle$ to $\vert {\tilde h}_{4,1} \rangle$
($1.502..\to 1.172..$),
$\vert {\tilde h}_{1,4}\rangle$ to $\vert {\tilde h}_{1,1} \rangle$
($1.256..\to 0.$)
\item $\vert {\tilde h}_{1,4}\rangle$ to $\vert {\tilde h}_{4,1}\rangle$
($1.256..\to 1.172..$)
\end{enumerate}
\end{enumerate}

\end{itemize}

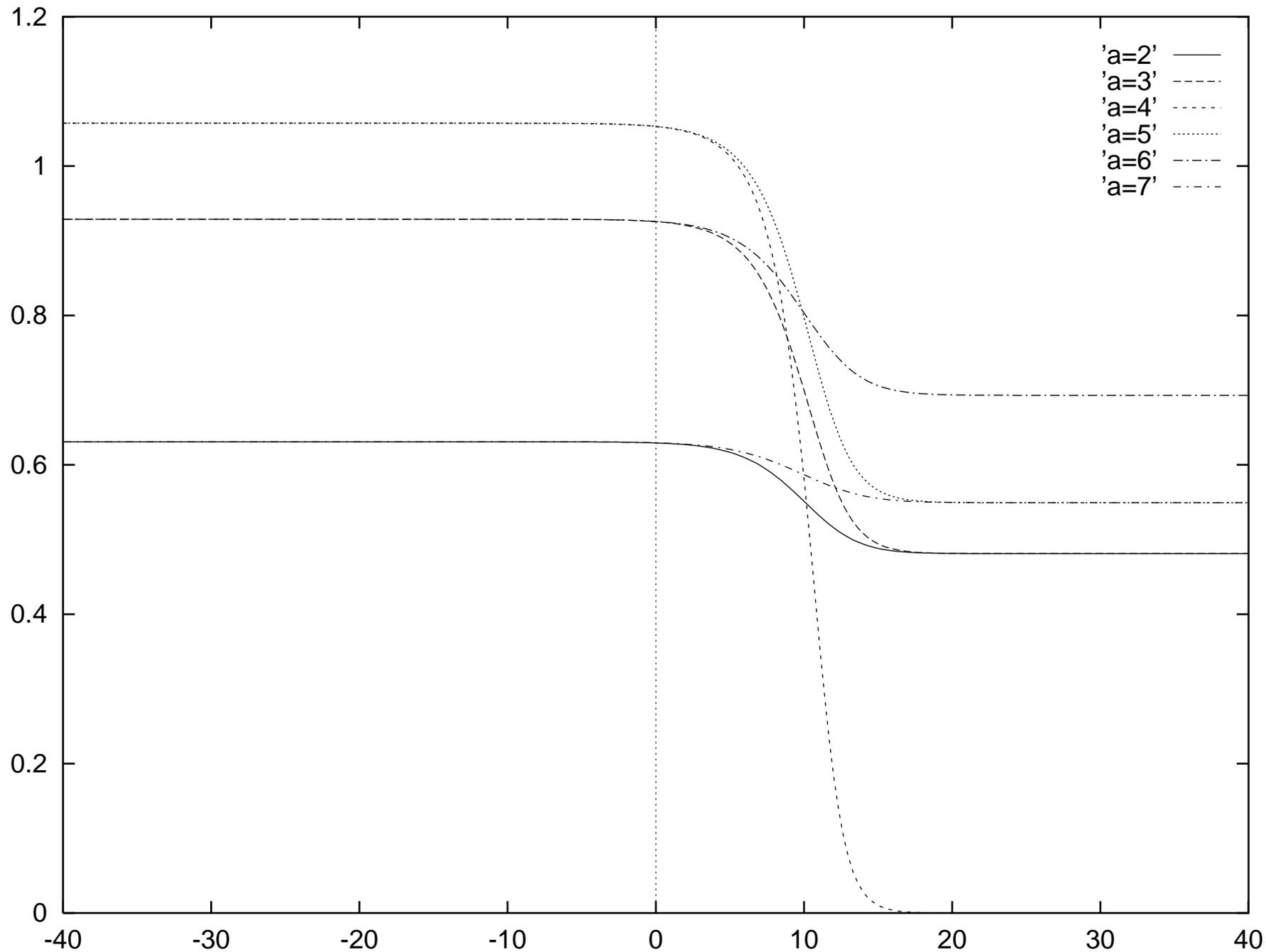
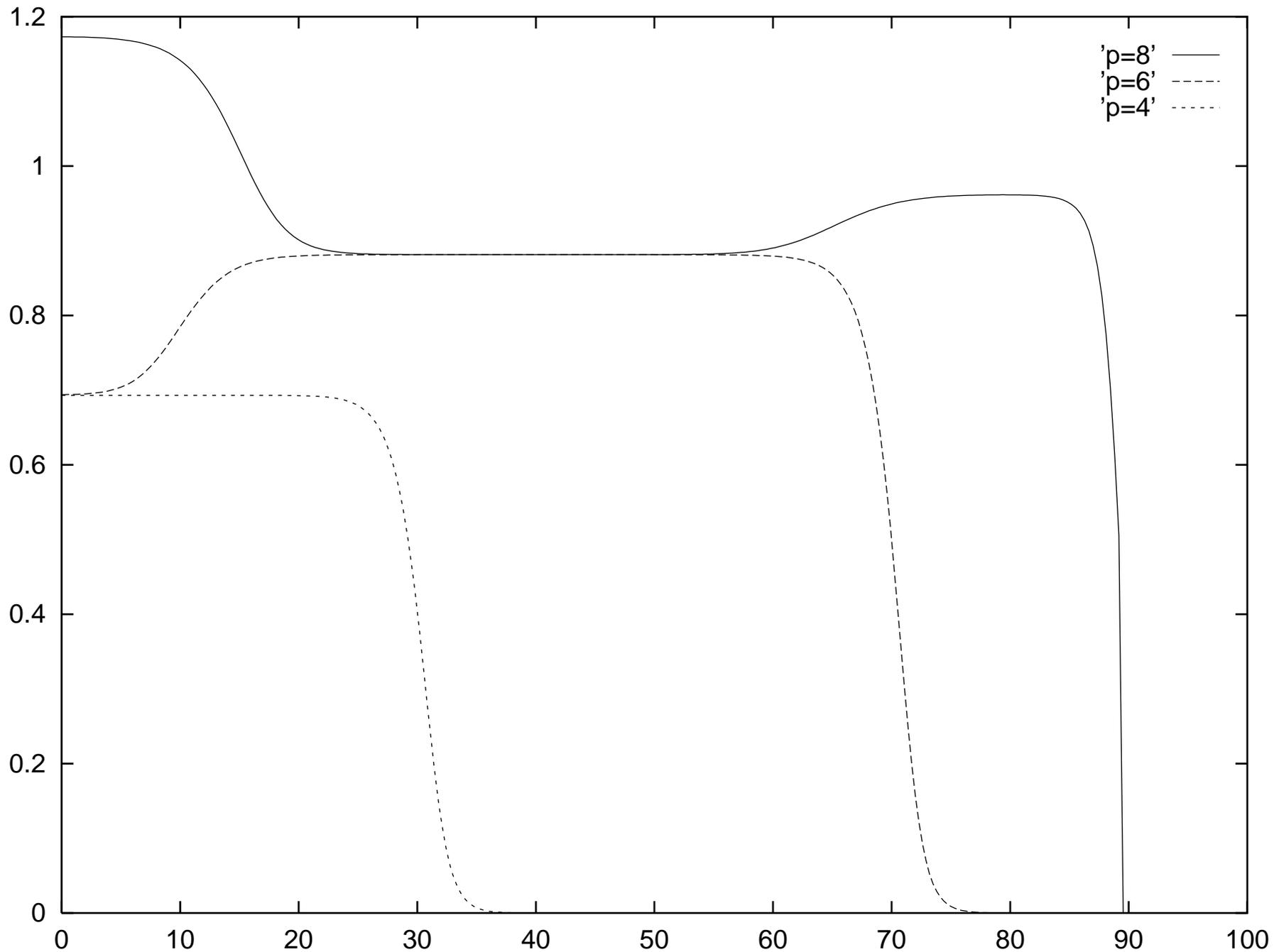
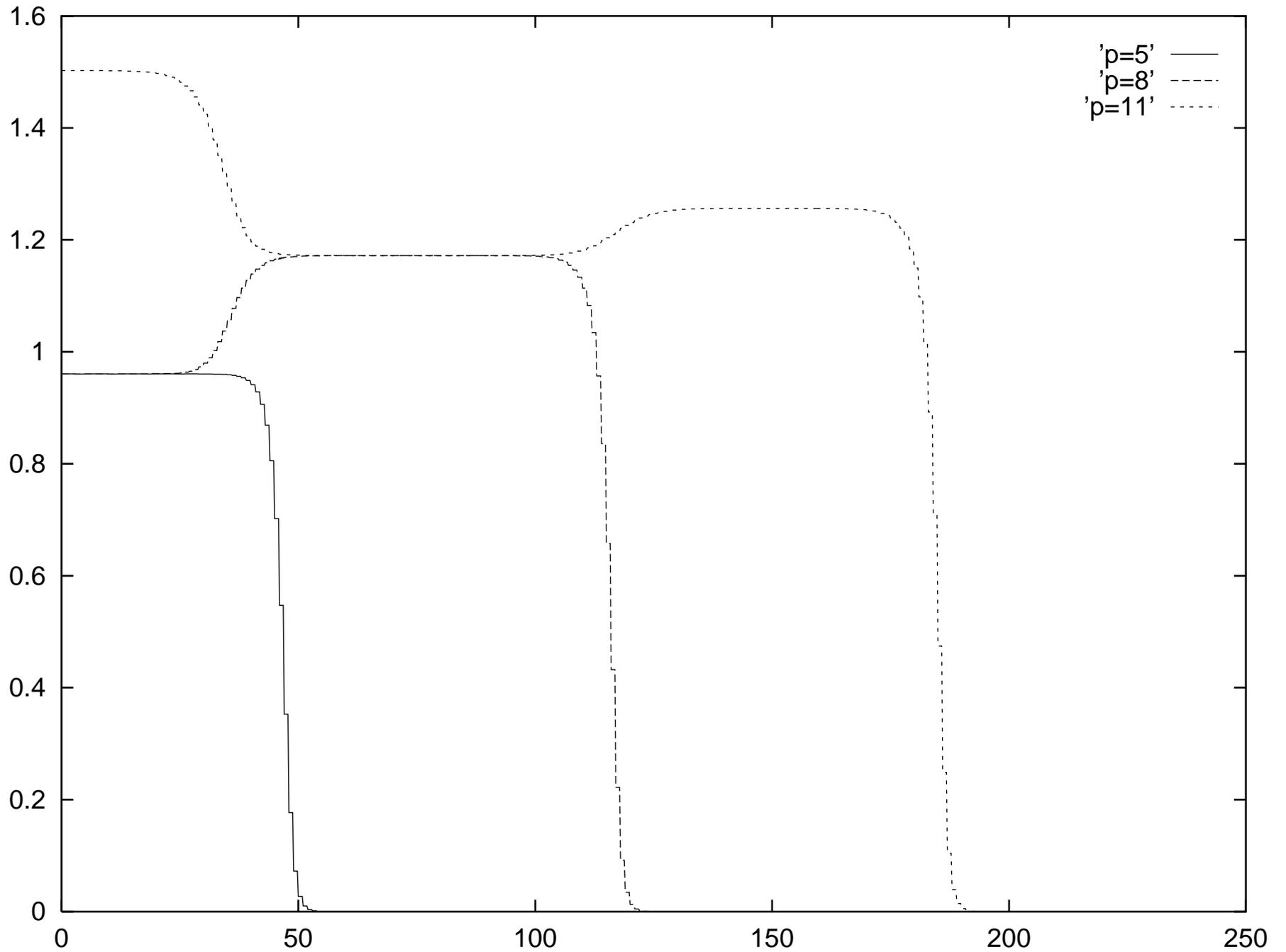
\end{document}